\documentclass[twocolumn,superscriptaddress,nobibnotes,footinbib,floatfix,aps,prb,citeautoscript,reprint]{revtex4-2}

\usepackage{graphicx}
\usepackage{epsfig}
\usepackage{subfigure}
\usepackage{color}
\usepackage{latexsym}
\usepackage{amsmath,bm,multirow}
\usepackage{amsfonts}
\usepackage{dcolumn}           
\usepackage{natbib}
\usepackage[colorlinks,linkcolor=blue,citecolor=green]{hyperref}
\usepackage{physics}

\usepackage[draft]{changes} % final OR draft
\usepackage[normalem]{ulem}
\definechangesauthor[name={yi}, color=blue]{yi}
\newcommand{\kl}{$\kappa_{\rm{L}}$}

\usepackage[draft]{changes} % final OR draft
\usepackage[normalem]{ulem}
\definechangesauthor[name={Maria}, color=red]{Maria}
\definechangesauthor[name={Yi}, color=blue]{Yi}
\usepackage{soul}
\begin{document}

%%%%%%%%%%%%%%%%%%%%%%%%%%%%%%%%%%%%%%%%%%%%%%%%%%%%%%%%%%%%%%%%%%%%%
%% The document title should be given as usual. Some journals require
%% a running title from the author: this should be supplied as an
%% optional argument to \title.
%%%%%%%%%%%%%%%%%%%%%%%%%%%%%%%%%%%%%%%%%%%%%%%%%%%%%%%%%%%%%%%%%%%%%
%\title[An \textsf{achemso} demo] {A demonstration of the \textsf{achemso} \LaTeX\ class\footnote{A footnote for the title}}
%\title{Data-Driven Understanding of Temperature-Dependent Phonons}
%\title{A Scalable Data-Driven Framework for Predicting Temperature-Dependent Phonons Across Materials}
\title{Data-Driven Exploration and Insights into Temperature-Dependent Phonons in Inorganic Materials}

\author{Huiju Lee\textsuperscript{\textdagger}}
\affiliation{Department of Mechanical and Materials Engineering, Portland State University, Portland, OR 97201, USA}
\thanks{These authors contributed equally to this work.}

%\alsoaffiliation{These authors contributed equally to this work.}

\author{Zhi Li\textsuperscript{\textdagger}}
\affiliation{Department of Materials Science and Engineering, Northwestern University, Evanston, IL 60208, USA}
\thanks{These authors contributed equally to this work.}

%\alsoaffiliation{These authors contributed equally to this work.}

\author{Jiangang He}
\affiliation{School of Mathematics and Physics, University of Science and Technology Beijing, Beijing 100083, China}

\author{Yi Xia}
\email{yimaverickxia@gmail.com; yxia@pdx.edu}
\affiliation{Department of Mechanical and Materials Engineering, Portland State University, Portland, OR 97201, USA}

\date{\today}

%%%%%%%%%%%%%%%%%%%%%%%%%%%%%%%%%%%%%%%%%%%%%%%%%%%%%%%%%%%%%%%%%%%%%
%% The abstract environment will automatically gobble the contents
%% if an abstract is not used by the target journal.
%%%%%%%%%%%%%%%%%%%%%%%%%%%%%%%%%%%%%%%%%%%%%%%%%%%%%%%%%%%%%%%%%%%%%
\begin{abstract}
  Phonons, quantized vibrations of the atomic lattice, are fundamental to understanding thermal transport, structural stability, and phase behavior in crystalline solids. Despite advances in computational materials science, most predictions of vibrational properties in large materials databases rely on the harmonic approximation and overlook crucial temperature-dependent anharmonic effects. Here, we present a scalable computational framework that combines machine learning interatomic potentials, anharmonic lattice dynamics, and high-throughput calculations to investigate temperature-dependent phonons across thousands of materials. By fine-tuning the universal M3GNet interatomic potential using high-quality phonon data, we improve phonon prediction accuracy by a factor of four while preserving computational efficiency. Integrating this refined model into a high-throughput implementation of the stochastic self-consistent harmonic approximation, we compute temperature-dependent phonons for 4,669 inorganic compounds. Our analysis identifies systematic elemental and structural trends governing anharmonic phonon renormalization, with particularly strong manifestations in alkali metals, perovskite-derived frameworks, and related systems. Machine learning models trained on this dataset identify key atomic-scale features driving strong anharmonicity, including weak bonding, large atomic radii, and specific coordination motifs. First-principles validation confirms that anharmonic effects can dramatically alter lattice thermal conductivity by factors of two to four in some materials. This work establishes a robust and efficient data-driven approach for predicting finite-temperature phonon behavior, offering new pathways for the design and discovery of materials with tailored thermal and vibrational properties.
\end{abstract}

\maketitle

%%%%%%%%%%%%%%%%%%%%%%%%%%%%%%%%%%%%%%%%%%%%%%%%%%%%%%%%%%%%%%%%%%%%%
%% Start the main part of the manuscript here.
%%%%%%%%%%%%%%%%%%%%%%%%%%%%%%%%%%%%%%%%%%%%%%%%%%%%%%%%%%%%%%%%%%%%%

%---------------------------------------------------------
%%%%%%%%%%%%%%%%%%%%%%%%%%%%%%%%%%%%%%%%%%%%%%%%%%%%%%%%%%
%---------------------------------------------------------
\section{Introduction}
All materials are made up of atoms, which vibrate ubiquitously, even at absolute zero temperature, due to quantum effects. When present in periodic solids, these vibrations can be quantized as quasiparticles, known as phonons~\cite{ashcroft1976solid}. The study of phonons is an integral part of solid-state physics and materials science, as they play an essential role in many physical properties of materials, including thermodynamical and transport phenomena~\cite{wallace1998thermodynamics}. To study phonons and their thermodynamic properties, modern simulations have relied on the foundational concept of harmonic approximation, which involves creating a lattice dynamics model using a second-order Taylor expansion of the Born-Oppenheimer potential energy surface. Despite its simplicity, the harmonic approximation has enjoyed immense success in explaining numerous phenomena observed in solid-state materials. This includes phenomena like phonon excitation spectra~\cite{parlinski1997first}, phase stability of various materials~\cite{grimvall2012lattice}, elastic properties~\cite{boer2023elasticity}, and zero-point motion~\cite{allen1994zero}. Consequently, it has naturally become one of the cornerstone tools in the field of condensed matter physics and materials science.

However, it is important to note that the early truncation of Taylor expansion of the Born-Oppenheimer potential energy surface  can significantly impact the vibrational spectrum. This is because the higher-order terms in the Taylor expansion, known as anharmonic terms, play a crucial role in determining phonon dynamics at finite temperatures. As long as the phonon quasiparticle picture applies, the temperature-dependent behavior of phonons can be categorized into two parts, namely, phonon frequency shifts and broadening. The former is usually referred to as anharmonic phonon renormalization (APRN), which shifts the phonon frequencies relative to the harmonic approximation, while the latter characterizes the finite lifetimes of phonon modes~\cite{wallace1998thermodynamics}. Probably, two of the most striking examples that anharmonic terms reflect themselves are the dynamical stabilization of crystal structures with imaginary phonon frequencies at 0~K and the finite value of lattice thermal conductivity ($\kappa_{\rm L}$)~\cite{cowley1963lattice}. To account for temperature-dependent phonons (TDPH) arising from APRN beyond harmonic/quasiharmonic approximation, recent advances in anharmonic lattice dynamics (ALD) simulations have centered on accounting for anharmonic terms using either perturbation theory or non-perturbative approaches. Specific realizations include the Stochastic Self-Consistent Harmonic Approximation (SSCHA)~\cite{SSCHA}, the Temperature-Dependent Effective Potential (TDEP)~\cite{TDEP2011}, and the Self-Consistent Phonon Theory (SCPH)~\cite{SCPH2015}, among others~\cite{pbte2018}. By integrating these approaches with first-principles calculations based on density functional theory (DFT)~\cite{dft1, dft2}, recent studies have unveiled significant insights into materials' thermodynamic and thermal transport properties relying on TDPH~\cite{li2015orbitally,budai2014metallization,li2014phonon,kim2018nuclear,errea2020quantum,kim2023chiral}. For example, they have revealed the nontrivial role of quartic anharmonicity in shaping heat transfer in semiconductors~\cite{feng2017four, rczbprx, Tadano2022}, the microscopic mechanisms of abnormal glasslike $\kappa_{\rm L}$ in crystalline compounds~\cite{tadano2015impact,tetrahedriteprl}, the quantum paraelectricity stabilized by anharmonic effects~\cite{VerdiQuantum}, and the temperature-dependent topological phonons~\cite{bopeng2020}, to name a few.

Despite these encouraging advances, a comprehensive understanding of TDPH across the periodic table remains elusive. Modern materials design necessitates systematic knowledge of the general trends governing the elemental and structural dependence of TDPH across diverse material classes. Such understanding is particularly crucial for identifying materials with exceptional thermal management capabilities for electronics and energy applications~\cite{Bell1457}, predicting structural phase transitions and stability regions with enhanced accuracy~\cite{bianco2017second}, and designing materials with tailored phonon-mediated properties including superconductivity~\cite{errea2020quantum} and thermoelectricity~\cite{tetrahedriteprl}. However, the formidable computational cost associated with ALD calculations has precluded such systematic investigations, leaving this fundamental understanding largely unexplored.

To address this challenge, we propose a data-driven approach to exploring TDPH at an unprecedented scale. Our approach integrates an efficient algorithm to quantify APRN with high-throughput calculations using a fine-tuned foundation model for machine learning interatomic potentials. This methodology achieves computational efficiency orders of magnitude superior to traditional approaches relying exclusively on DFT calculations, thereby enabling explicit screening of TDPH across thousands of inorganic compounds. Through statistical analysis of our extensive dataset, we reveal distinct trends and unique behaviors of APRN across the elemental distribution and structural space groups. Furthermore, we employ machine learning models to identify critical atomic environments that underlie strong APRN, providing fundamental chemical and physical insights into TDPH mechanisms. Finally, we demonstrate the practical utility of our approach through materials screening and rigorous first-principles validation of downselected materials exhibiting strong APRN, highlighting their impact on $\kappa_{\rm L}$. The remarkable efficiency of our framework significantly enhances the potential for materials discovery and design applications with tailored vibrational properties.

%--------------------------------------------------------------------------------------------------------------------------
\begin{figure*}[htp]
%\begin{figure*}
	\includegraphics[width = 1.0\linewidth]{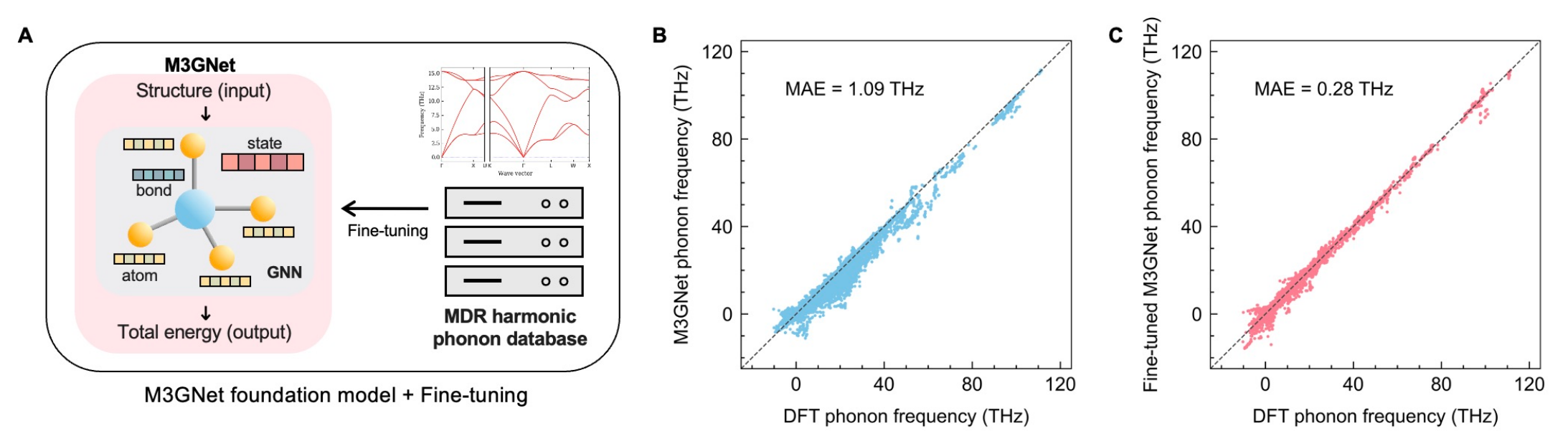}
	\caption{ %\small
    (A) Illustration of the M3GNet model, based on a graph neural network (GNN) and fine-tuned with the MDR harmonic phonon database to improve phonon property predictions. (B) Comparisons of phonon frequencies predicted by the original M3GNet foundation model and those calculated using DFT. (C) Comparisons of phonon frequencies predicted by the fine-tuned M3GNet foundation model and DFT results.
	}
	\label{fig:m3gnet}
\end{figure*}
%--------------------------------------------------------------------------------------------------------------------------

%---------------------------------------------------------
%%%%%%%%%%%%%%%%%%%%%%%%%%%%%%%%%%%%%%%%%%%%%%%%%%%%%%%%%%
%---------------------------------------------------------
\section{Results and discussion}

%---------------------------------------------------------
%---------------------------------------------------------
\subsection{Fine-tuned M3GNet foundation model for enhanced phonon predictions}

Machine learning interatomic potentials (MLIPs) have revolutionized computational materials science by enabling atomistic simulations with near-quantum mechanical accuracy at a fraction of the computational cost~\cite{unke2021machine}. Foundation models, such as Materials Graph Network with three-body interactions (M3GNet)~\cite{Chen2022}, Multi-Atomic Cluster Expansion (MACE)~\cite{batatia2023foundation}, and Crystal Hamiltonian Graph neural Network (CHGNet)~\cite{Deng2023}, offer pre-trained models with universal applicability across different materials systems, eliminating the need for extensive training data generation for each new system. In this study, we employ the M3GNet model as the backbone for computing forces and energies due to its proven reliability, computational efficiency, and extensive validation across diverse materials systems. However, M3GNet exhibits notable limitations in predicting phonon properties, systematically underestimating the phonon density of states compared to DFT calculations, with a large mean absolute error of 44.2 cm$^{-1}$ (1.32 Thz)~\cite{Chen2022}.

To address this known limitation of M3GNet in phonon predictions, we fine-tune the M3GNet foundation model using high-quality displacement-force data generated from the Materials Data Repository (MDR) phonon calculation database~\cite{togo_database}, as illustrated in Fig.~\ref{fig:m3gnet}A. Specifically, our approach leverages randomly generated small atomic displacements, consistent with the harmonic approximation underlying the MDR phonon database, to construct a comprehensive force-displacement dataset by performing tensor product between harmonic force constants and displacements. Such a dataset encompasses a wide range of atomic environments and vibrational modes across different crystal structures and chemical compositions. We refer readers to Methods for more details on the fine-tuning process. This strategic fine-tuning yields dramatic improvements in phonon prediction accuracy, reducing the phonon frequency mean absolute error (MAE) from 1.09 THz (Fig.~\ref{fig:m3gnet}B) to 0.28 THz (Fig.~\ref{fig:m3gnet}C), a fourfold reduction in MAE compared to the original M3GNet model. The improvement is particularly notable as it significantly reduces the systematic softening of potential energy surfaces observed in several universal MLIP models~\cite{deng2025systematic}, including M3GNet and CHGNet~\cite{Deng2023}. In M3GNet, phonon frequencies are consistently underestimated relative to DFT calculations, largely due to the use of a training dataset constructed with low-convergence DFT criteria and Perdew-Burke-Ernzerhof (PBE) exchange-correlation functiional~\cite{PBE} , in contrast to the MDR phonon database, which employs tighter convergence criteria and PBE for solids (PBEsol)~\cite{Perdew2008}. The fine-tuned model successfully mitigates this inherent bias while maintaining M3GNet's computational efficiency and broad chemical space coverage, achieving significantly enhanced reliability for vibrational property predictions.

Although our fine-tuning procedure is based solely on displacement-force data from harmonic phonon calculations, the enhanced M3GNet model retains its intrinsic ability to capture anharmonic effects through its underlying neural network architecture. We posit that this fine-tuning not only refines harmonic interactions but also implicitly adjusts anharmonic components of the potential energy surface, as the neural network internalizes improved atomic interaction patterns from high-quality training data that generalize across the full potential energy landscape. Nevertheless, we emphasize that we intend not to fully replace DFT calculations with M3GNet model. Instead, we utilize the fine-tuned M3GNet model primarily for rapid screening of TDPH within MDR phonon database and for analyzing the underlying chemical and structural trends. Final validation of the results on TDPH is performed using direct DFT calculations combined with rigorous ALD simulations.

%---------------------------------------------------------
%---------------------------------------------------------
\subsection{Streamlined calculation of anharmonic phonon renormalization}

%------------------------------------------------------------------------------------------------------
\begin{figure*}[htp]
	\centering
	\includegraphics[width = 1.0\linewidth]{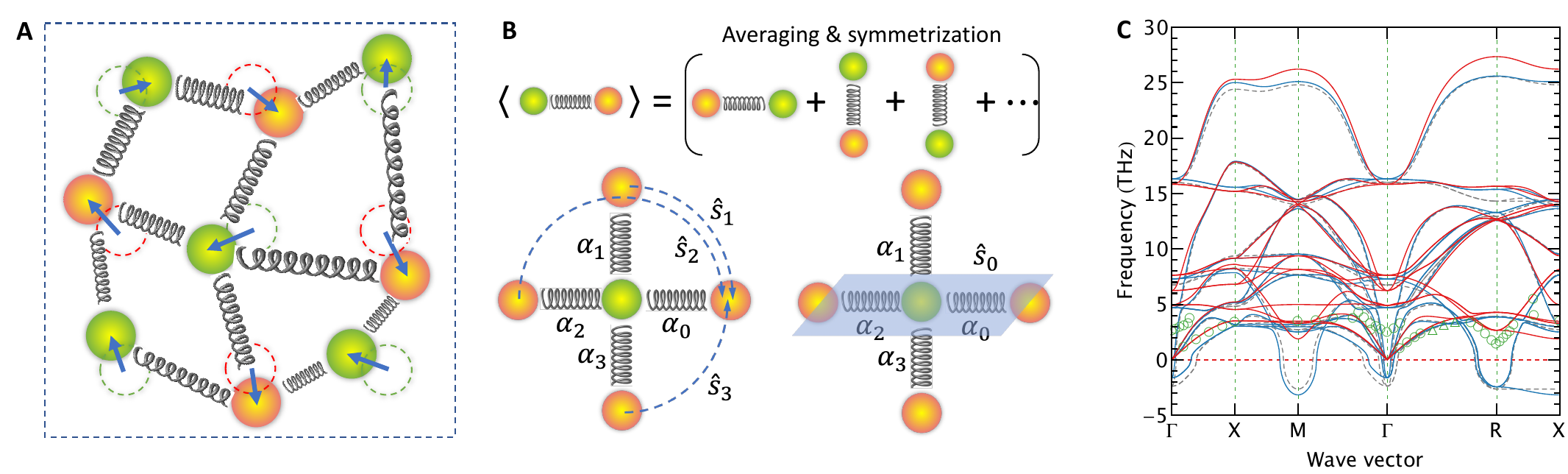}
	\caption{%\small
    Overview of anharmonic phonon renormalization (APRN) calculation and validation. (A) Schematic illustrating configuration-dependent interatomic force constants (represented by springs) between atoms (solid circles) displaced from their equilibrium positions (dashed circles). (B) Schematic showing the symmetrization of these force constants using space group symmetry operations, wherein $\alpha$ denotes a pair interaction and $\hat{S}$ indicate a symmetry operator ($\hat{S}_{1/2/3}$ for rotations and $\hat{S}_{o}$ for reflection). (C) Validation of the APRN approach using cubic SrTiO$_{3}$ as a test case. The phonon dispersions are shown as calculated from DFT (gray dashed lines), fine-tuned M3GNet harmonic approximation (blue solid lines), and renormalized calculations using fine-tuned M3GNet model at 300~K (red solid lines). Green circles represent experimental measurements at 300~K~\cite{stirling1972neutron,cowley1969relationship}.
	}
	\label{fig:tdifc}
\end{figure*}
%------------------------------------------------------------------------------------------------------

With this enhanced foundation model established, we now turn to calculating TDPH. Among available approaches, including TDEP~\cite{TDEP2011}, SCPH~\cite{SCPH2015}, and SSCHA~\cite{SSCHA}, we employed SSCHA to enable efficient high-throughput screening of TDPH. This choice is motivated by the computational limitations of alternative methods: TDEP typically relies on direct molecular dynamics simulations, while SCPH calculations require many-body anharmonic force constants, both of which demand substantial computational cost and complex convergence tests. To enable efficient computation of TDPH across large materials databases, we implemented a streamlined variant of SSCHA that significantly reduces computational overhead while maintaining sufficient accuracy for qualitative materials screening.

The SSCHA method relies on minimizing the Gibbs-Bogoliubov free energy to obtain effective harmonic force constants at finite temperatures $\Phi^{\rm SSCHA}_{i\alpha, j\beta}(T)$~\cite{SSCHA},
\begin{equation}
\label{eq:scha}
    \Phi^{\rm SSCHA}_{i\alpha, j\beta}(T) = 
    \biggl \langle \frac{\partial^2 V(\mathbf{R})}{\partial R_{i\alpha} \partial R_{j\beta} } \biggr \rangle_{T} = 
    \biggl \langle \Phi^{\rm HA}_{i\alpha, j\beta} (\mathbf{R}) \biggr \rangle_{T},
\end{equation}
wherein $V$ is the interatomic potential, $\mathbf{R}$ represents atomic positions (not necessarily at equilibrium), $i/j$ and $\alpha/\beta$ denote atom and Cartesian indices, respectively. In Eq.~[\ref{eq:scha}], $\langle \rangle$ indicates an ensemble average, meaning that $\Phi^{\rm SCHA}_{i\alpha, j\beta}(T)$ is obtained by averaging instantaneous harmonic force constants $\Phi^{\rm HA}_{i\alpha, j\beta} (\mathbf{R})$, calculated as second order derivatives of $V$ for specific atomic configurations $\mathbf{R}$, as illustrated in Fig.~\ref{fig:tdifc}A. Within the SSCHA framework, atomic displacements in the temperature-dependent configurations are sampled from a quantum covariance matrix $\Sigma_{u_{\mathbf{a}}, u_{\mathbf{b}}}$ \cite{Errea2014,Roekeghem2016,pbte2018}
	\begin{equation}\label{eq:covar}
	\Sigma_{u_{\mathbf{a}}, u_{\mathbf{b}}} = \frac{\hbar}{2\sqrt{m_{a}m_{b}}} \sum_{\lambda} \frac{\left(1+2n^{0}_{\lambda}\right) }{\omega_{\lambda}} e^{\lambda}_{\mathbf{a}} e^{\lambda\ast}_{\mathbf{b}},
	\end{equation}
where $n_{\lambda}^{0}$ and $\omega_{\lambda}$ are Bose-Einstein distribution function and vibrational frequency of phonon mode $\lambda$, respectively, while $m_{a}$ and $e^{\lambda}_{\mathbf{a}}$ are atomic mass and phonon eigenvector projected on atom $a$, respectively. The primary computational bottleneck of SSCHA arises from achieving convergence through evaluating configuration-dependent $\Phi^{\rm HA}_{i\alpha, j\beta} (\mathbf{R})$ and self-consistence by iterating between Eq.~[\ref{eq:scha}] and Eq.~[\ref{eq:covar}]. 
Our streamlined implementation of SSCHA adopts two key strategies. First, we perform single-shot calculations without self-consistency by utilizing $\Phi^{\rm HA}_{i\alpha, j\beta} (\mathbf{R})$ obtained from equilibrium structures in Eq.~[\ref{eq:covar}]. This is conceptually similar to the $G_0W_0$ approximation~\cite{hedin1965new,hybertsen1986electron}, which provides a one-shot correction to the Kohn-Sham eigenvalues obtained from DFT. This approximation captures the first-order correction and thereby enables a semiquantitative description of APRN. Second, instead of brute-force computation of the ensemble average in Eq.~[\ref{eq:scha}], we leverage space group symmetry to map symmetrically equivalent pair interactions, allowing us to obtain ensemble averaged force constants more efficiently, as illustrated in Fig.~\ref{fig:tdifc}B. Adopting these two strategies significantly improves computational efficiency while maintaining reasonable accuracy.

To demonstrate the efficacy of our streamlined SSCHA implementation integrated with the fine-tuned M3GNet model, we present a validation study using cubic perovskite SrTiO$_3$ as an exemplar system. Fig.~\ref{fig:tdifc}C compares harmonic phonons from DFT calculations and from our fine-tuned M3GNet, respectively, as well as TDPH at 300~K obtained using our SSCHA implementation, and available experimental measurements~\cite{stirling1972neutron,cowley1969relationship}. The excellent agreement between DFT and machine learning harmonic phonons further validates the effectiveness of our fine-tuning approach, while the close correspondence between experimental measurement and our TDPH predictions demonstrates the accuracy of the integrated methodology. Remarkably, these high-quality TDPH results were achieved using only a single 2$\times$2$\times$2 displaced supercell (40 atoms) throughout the calculation, highlighting the exceptional computational efficiency enabled by our symmetry-averaging approach. This dramatic reduction in computational requirements while maintaining reasonable accuracy, compared to first-principles approaches (e.g., SCPH) that typically require a much larger number of supercells calculated by DFT~\cite{SCPH2015}, underscores the practical utility of our framework for large-scale materials screening. 

We note that, strictly speaking, the phonons obtained from our streamlined SSCHA approach correspond to auxiliary phonons~\cite{SSCHA} -- analogous to the self-consistent phonons in many-body diagrammatic perturbation theory~\cite{mahan2000many} -- rather than direct physical observables of lattice vibrations. Nevertheless, these auxiliary modes effectively capture the temperature-dependent trends exhibited by experimentally measured phonons, as demonstrated in the case of SrTiO$_3$ above. This reinforces the reliability of our methodology for reproducing realistic anharmonic behavior while maintaining its computational efficiency.

%---------------------------------------------------------
%---------------------------------------------------------
\subsection{Elemental and structural trends in anharmonic behavior from large-scale screening}

%-----------------------------------------------------
\begin{figure*}[htp]
%\begin{figure*}
	\includegraphics[width = 1.0\linewidth]{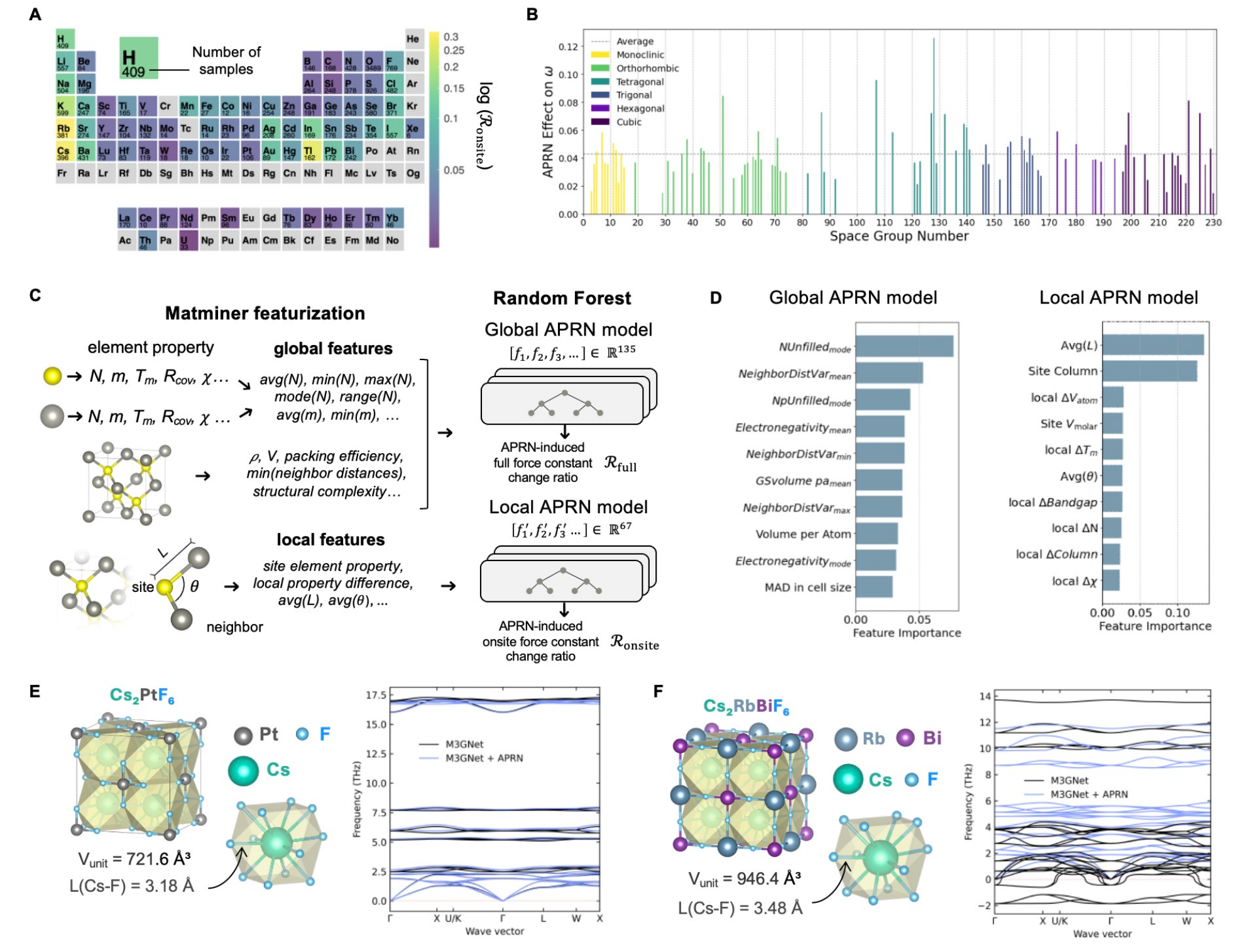}
	\caption{ %\small
    (A) Heatmap of the APRN-induced percentage change of onsite force constants ($\mathcal{R}_{\text{onsite}}$) across the periodic table, plotted on a logarithmic scale. Brighter colors indicate stronger APRN effects, highlighting element-wise trends in temperature-dependent phonons. 
    (B) Bar plot showing the average APRN effect on phonon frequencies for each space group number. The colors represent different crystal systems, and the dashed horizontal line indicates the overall average across all materials. 
    (C) Illustration of the Matminer featurization process and random forest model training. Global features (structural and compositional properties) are used to predict $\mathcal{R}_{\text{full}}$, while local features (site environments) are used to predict $\mathcal{R}_{\text{onsite}}$. These two features are used to train two separate models: global APRN model and local APRN model.
    (D) Feature importance rankings for the top 10 features in the global APRN model (left) and local APRN model (right).
    (E)-(F) Case studies of materials with weak vs. strong APRN effects, demonstrating how local structural properties such as bond length ($L$) and unit cell volume ($V_\text{unit}$) affect phonon renormalization. The phonon dispersions compare the harmonic predictions (M3GNet, black) with the APRN-corrected predictions at 300~K (M3GNet + APRN, blue).
	}
	\label{fig:random-forest}
\end{figure*}
%-----------------------------------------------------

Having validated our computational approach, we next screened TDPH across a large materials database to achieve a comprehensive understanding of the structural and chemical dependence of TDPH. We selected the MDR phonon database for this comprehensive study, leveraging our fine-tuned M3GNet model that was specifically optimized for this dataset. To ensure computational feasibility while maintaining broad coverage, we further downselected materials containing fewer than 20 atoms per unit cell, yielding 4,669 compounds (Fig.~\ref{fig:prototypes}A) for which we performed streamlined TDPH calculations at 300~K.

To systematically quantify APRN effects at 300 K across this comprehensive dataset, we developed a comparative framework that analyzes model-predicted force constants before and after APRN using our fine-tuned M3GNet foundation model. This analysis employs two complementary metrics designed to capture different aspects of anharmonic behavior: (1) the APRN-induced full force constant change ratio $\mathcal{R}_{\text{full}}$, which measures global changes across the entire force constant matrix (with a dimension of $3N_{\rm sc} \times 3N_{\rm sc}$, and $N_{\rm sc}$ denotes the number of atoms in a supercell), and (2) the APRN-induced onsite force constant change ratio $\mathcal{R}_{\text{onsite}}$ (with a dimension of $3 \times 3$), which focuses on local atomic environment effects. This dual approach allows us to distinguish between materials with uniform anharmonicity and those with localized effects.

The quantity $\mathcal{R}_{\text{full}}$ is defined as the mean relative change across all components of the force constant matrix resulting from APRN:
\begin{equation}
\label{eq:r_fc}
    \mathcal{R}_{\text{full}} = \text{Mean} \left( \frac{\left| \Phi^{\text{APRN}}_{ij} - \Phi^{\text{orig}}_{ij} \right|}{\left( \left| \Phi^{\text{orig}}_{ij} \right| + \left| \Phi^{\text{APRN}}_{ij} \right| \right)/2} \right)
\end{equation}
where \(\Phi^{\text{orig}}_{ij}\) and \(\Phi^{\text{APRN}}_{ij}\) represent the force constant values between atom pair $i$ and $j$ before APRN and after the APRN correction at 300 K, respectively. The mean value is obtained by averaging all atom pairs within a given supercell for phonon calculation. We adopt the arithmetic mean as the denominator to ensure numerical stability. This choice is motivated by cases where force constant values are close to zero, as it avoids division by small values that could produce excessively large or misleading percentage changes. While $\mathcal{R}_{\text{full}}$ reflects a structure-wide view of the APRN effect due to the use of the full force constants matrix, we introduce another quantity, $\mathcal{R}_{\text{onsite}}$, to capture changes at the individual atomic environment. This distinction is important because, within a given material, atoms can exhibit varying degrees of APRN depending on their local bonding environments. Consequently, a single material can possess multiple $\mathcal{R}_{\text{onsite}}$ values corresponding to each type of atomic environment. $\mathcal{R}_{\text{onsite}}$ quantifies the relative changes in the trace of the onsite force constant matrix:
\begin{equation}
\label{eq:r_ofc}
    \mathcal{R}_{\text{onsite}} = \left|\frac{\text{tr}(\Phi^{\text{APRN}}_{ii}) - \text{tr}(\Phi^{\text{orig}}_{ii})}{\text{tr}(\Phi^{\text{orig}}_{ii})}\right|
\end{equation}
where \(\text{tr}(\Phi^{\text{orig}}_{ii})\) and \(\text{tr}(\Phi^{\text{APRN}}_{ii})\) denote the trace of the onsite force constant matrix before APRN and after the APRN correction at 300 K, respectively.

We first analyze the element-wise distribution of APRN effects quantified by $\mathcal{R}_{\text{onsite}}$ across the periodic table (Fig.~\ref{fig:random-forest}A). Certain elements, such as Rb, Cs, and Tl, exhibit significantly stronger TDPH compared to others. This suggests that alkali metals and some elements with specific bonding characteristics are prone to exhibit stronger TDPH at 300 K. Alkali elements typically have large atomic radii and low electronegativities, leading to weaker bonding and enhanced anharmonicity~\cite{straus2020understanding,laven2025unraveling}. Tl, due to its heavy atomic mass, contributes to low-frequency phonon modes, which are inherently more prone to anharmonic effects~\cite{mukhopadhyay2018two,zeng2024pushing}. Furthermore, in certain crystal structures, such as cage- or chain-like structures, Tl atoms often occupy weakly bonded or oversized sites, leading to rattling motions that enhance anharmonicity~\cite{yixiaTVS,dutta2019ultralow,pal2021microscopic,jana2017intrinsic}. We note that hydrogen also exhibits considerable APRN, largely due to strong quantum effects stemming from its extremely light mass~\cite{errea2014anharmonic,ceriotti2016nuclear}, although it does not rank among the strongest elements shown above. This is because the present comparison assesses the strength of APRN relative to the harmonic approximation, under which hydrogen typically forms strong bonds and consequently displays high-frequency phonon modes.

In the analysis by space group, the overall APRN effect is determined by first calculating a single value that captures the total change across all phonon frequencies. For each material, this value is computed using L2 norm as follows: $\lVert \boldsymbol{\omega}_{\mathrm{APRN}} - \boldsymbol{\omega}_{\mathrm{orig}} \rVert_2 / \lVert \boldsymbol{\omega}_{\mathrm{orig}} \rVert_2$, where $\boldsymbol{\omega}_{\mathrm{orig}}$ and $\boldsymbol{\omega}_{\mathrm{APRN}}$ are the vectors of original harmonic (0~K) and the vector of APRN (300~K) phonon frequencies, respectively, both predicted by the fine-tuned M3GNet model. The APRN effect reported for each space group is then obtained by averaging these values over all compounds belonging to that group. As shown in Fig.~\ref{fig:random-forest}B, significant APRN effects are particularly observed in tetragonal and cubic crystal systems. Notably, space groups 128, 107, 221, 87, and 225 exhibit strong APRN effects and are commonly found in perovskite-related or structurally similar materials. For instance, space group 128 typically corresponds to tetragonal double perovskites, space group 87 is related to distorted tetragonal perovskites, and space groups 221 and 255 are associated with cubic perovskite and double perovskite structures. Additionally, space group 107 includes structures derived from double perovskites, such as K$_2$NaTiOF$_5$, which share similarities with K$_2$NaAlF$_6$. The high APRN effects in these perovskite-related materials can often be attributed to structural instabilities at certain temperatures, such as octahedral tilting and the presence of rattling ions~\cite{da2015phase,Yang2017,hoffman2023understanding,klarbring2020anharmonicity}.

To systematically understand the underlying factors driving APRN effects, we build two distinct random forest models: a global APRN model and a local APRN model (Fig.~\ref{fig:random-forest}C) to learn $\mathcal{R}_{\text{full}}$ and $\mathcal{R}_{\text{onsite}}$, respectively. The global APRN model is trained using composition- and structure-based features derived from the Matminer package~\cite{ward2018matminer}. For composition-based features, statistical descriptors (e.g., average, minimum, maximum, mode, and range) of elemental properties, including Mendeleev number ($N$), atomic weight ($m$), covalent radius ($R_{cov}$), electronegativity ($\chi$), among others, are computed across the elements in each compound. For structure-based features, we include global structural descriptors such as density ($\rho$), unit cell volume ($V$), and packing efficiency. These features collectively capture the macroscopic attributes of a material that may influence $\mathcal{R}_{\text{full}}$. The local APRN model, in contrast, is designed to capture site-specific effects that contribute to APRN by analyzing the local environment around individual atoms. Here, a site refers to a specific atomic position within a material's crystal structure. The local environment of a site is defined by the bonding interactions and spatial arrangement of its neighboring atoms. The model utilizes features derived from each site and its neighbors, including elemental properties of the site atom, differences in properties, average bond length ($L$), and average bond angle ($\theta$) between the site and neighboring atoms. While the global model highlights broad compositional and structural trends influencing APRN, the local APRN model offers a complementary perspective by focusing on local atomic environments to learn the correlation between local features and $\mathcal{R}_{\text{onsite}}$. We refer the readers to Methods for more details on training random forest models.

Fig. \ref{fig:random-forest}D shows the top 10 most important features for both the global and local APRN models. In the global model, the most dominant feature is the mode of the number of unfilled orbitals (NUnfilled\textsubscript{mode}), followed by descriptors such as the mean neighbor distance variation (NeighborDisVar\textsubscript{mean}), the mode of unfilled p-orbitals (NpUnfilled\textsubscript{mode}), and the mean electronegativity. Here, the mode refers to the value that appears most frequently in a data set. In the local model, the average bond length (Avg($L$)) and the column number of the site atom in the periodic table are the top contributors, followed by the local difference in volume per atom ($\Delta V_\text{atom}$) and the molar volume of the site atom (site $V_\text{mole}$). These feature rankings reveal the physical origins of anharmonicity. Large atomic radii and weak bonding (reflected in bond length and volume features) create loose atomic environments prone to large-amplitude vibrations. Electronic configuration features (unfilled orbitals) indicate that partially filled electronic shells contribute to softer potential energy surfaces, enhancing anharmonic effects. We emphasize that these identified features represent general, data-driven trends rather than strict rules, and exceptions are expected. Moreover, the apparent feature importance is naturally limited by the structural and chemical diversity present in our dataset. Nonetheless, the strength of our model lies in offering an intuitive and computationally efficient framework for screening materials with strong APRN by linking these effects directly to local atomic environments.

Two representative materials showing weak and strong APRN effects are illustrated in Fig.~\ref{fig:random-forest}(E-F), respectively. These examples demonstrate how differences in the local environment, such as bond length and unit cell volume, can influence the degree of phonon renormalization. Both materials, Cs$_2$PtF$_6$ and Cs$_2$RbBiF$_6$, crystallize in cubic structures, where each Cs$^{1+}$ ion is bonded to twelve equivalent F$^{1-}$ ions, forming CsF$_{12}$ cuboctahedra. Despite this structural similarity, the two materials exhibit contrasting TDPH behaviors due to differences in their local environments. Cs$_2$RbBiF$_6$ has longer Cs-F bond lengths ($L_\text{Cs-F}$) and a larger unit cell volume ($V_\text{unit}$) compared to Cs$_2$PtF$_6$, resulting in a softer local environment. The phonon dispersion plots clearly reflect this difference. For Cs$_2$PtF$_6$, the harmonic prediction (M3GNet, black lines) and the prediction including anharmonic effects at 300~K (M3GNet + APRN, blue lines) are nearly identical. In contrast, Cs$_2$RbBiF$_6$ shows significant frequency shifts when comparing the M3GNet and M3GNet+APRN results, indicating strong TDPH.

%---------------------------------------------------------
%---------------------------------------------------------
\subsection{First-principles validation and thermal transport implications}
%---------------------------------------------------------
\begin{figure*}[htp]
	\includegraphics[width = 1.0\linewidth]{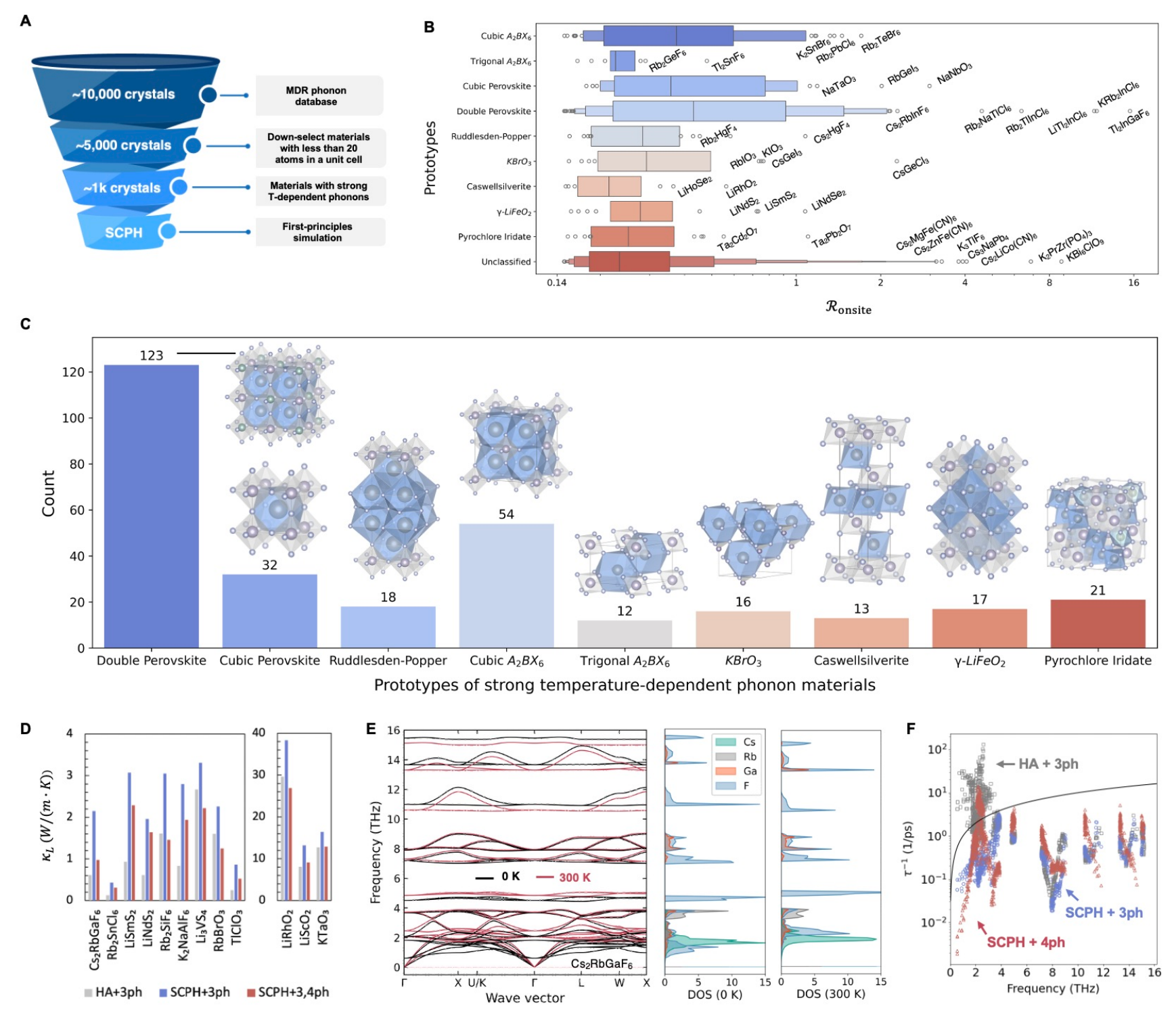}
	\caption{%\footnotesize
    (A) Schematic workflow for screening materials with strong temperature-dependent phonons (TDPH). First, for computational feasibility, we selected $\sim$5,000 materials with fewer than 20 atoms per unit cell from the MDR phonon database. This subset was then filtered to $\sim$1,200 candidate materials based on their APRN-induced onsite force constant change ratio $(\mathcal{R}_{\text{onsite}})$, using a threshold of $\mathcal{R}_{\text{onsite}} \geq 0.15$. This threshold corresponds to a 15\% change in onsite force constants due to the APRN effect. The screening was followed by validation with SCPH theory.
    (B) Boxen plot showing $\mathcal{R}_{\text{onsite}}$ values for various material prototypes within the 1,200 selected crystals. Top-ranked prototypes exhibiting strong APRN contributions are shown, providing insight into structural trends associated with significant temperature-dependent APRN effects. 
    (C) Distribution of material prototypes among the 1,200 screened materials, along with structural representations of key prototypes. The blue-colored polyhedral structures such as cuboctahedra and octahedra illustrate typical local atomic environments where onsite atoms exhibit strong temperature-dependent phonon behavior. 
    (D) Comparisons of lattice thermal conductivities (\kl) for 12 selected materials at 300 K, calculated using different levels of theory: harmonic approximation (without anharmonic renormalization) with three-phonon scattering (HA + 3ph), self-consistent phonon approximation (with anharmonic renormalization) with three-phonon scattering (SCPH + 3ph), and SCPH with both 3- and 4-phonon scattering (SCPH 3,4ph).
    (E) Phonon dispersion and phonon density of states (DOS) for Cs$_2$RbGaF$_6$ at 0 K (black) and 300 K (red), showing an APRN effect primarily driven by contributions from F$^-$ ions.
    (F) Phonon scattering rates ($\tau^{-1}$) of Cs$_2$RbGaF$_6$ computed using HA + 3ph, SCPH + 3ph, and SCPH + 4ph approximations. The solid black line represents the Mott-loffe-Regel limit, where the scattering rate equals the phonon frequency.}
	\label{fig:prototypes}
\end{figure*}
%---------------------------------------------------------

Having obtained a global understanding of TDPH behavior across the broad spectrum of materials from the MDR database, we move on to identify and downselect materials with strong APRN, followed by first-principles validation using SCPH theory (Fig.~\ref{fig:prototypes}A). From the MDR phonon database, we screened materials based on $\mathcal{R}_{\text{onsite}}$, using a threshold of 0.15, which yielded approximately 1,200 candidate materials. This criterion corresponds to a 15\% change in onsite force constants arising from APRN, indicating significant phonon renormalization at one or more atomic sites at finite temperatures. Figure~\ref{fig:prototypes}B presents the distribution of $\mathcal{R}_{\text{onsite}}$ values for the nine most frequently occurring structural prototypes among these strong SCPH candidate materials. The list of these materials, including their chemical formula and Materials Project IDs, is provided in the Supplementary Material. Prototype classification is based on the AFLOW Encyclopedia of Crystallographic Prototypes~\cite{mehl2017aflow, hicks2019aflow,  hicks2021aflow, eckert2024aflow}. In Fig.~\ref{fig:prototypes}B, each data point represents the highest $\mathcal{R}_{\text{onsite}}$ value for the given material. The analysis reveals that perovskite-related materials, such as double perovskite, cubic perovskite, Ruddlesden-Popper phases, and the cubic A$_2$BX$_6$ structure which is a vacancy-ordered double perovskite, tend to exhibit high $\mathcal{R}_{\text{onsite}}$. Other prototypes, such as trigonal A$_2$BX$_6$, KBrO$_3$, caswellsilverite (CrNaS$_2$), $\gamma$-LiFeO$_2$ (a promising lithium-ion battery anode material~\cite{guo2017first}), and pyrochlore iridates, also show noticeable APRN effects. Notably, $\gamma$-LiFeO$_2$ has a caswellsilverite-like structure, sharing similar octahedral coordination environments~\cite{MaterialsProject}.

Fig.~\ref{fig:prototypes}C shows the distribution of the top nine structural prototypes among approximately 1,200 screened materials. Double perovskite structures exhibit the highest average $\mathcal{R}_{\text{onsite}}$, followed by cubic A$_2$BX$_6$ and cubic perovskites. The inset structural illustrations provide representative examples of these prototypes, highlighting common local motifs such as cuboctahedra, octahedra, and anticuboctahedra, which are associated with high $\mathcal{R}_{\text{onsite}}$. In these insets, the blue-colored polyhedra indicate atomic environments showing strong APRN effects. Especially, cuboctahedra and anticuboctahedra are 12-fold coordination environments that often surround large cations, such as alkali metals. These sites are typically oversized and weakly bonded, enabling large amplitude rattling motions that enhance phonon–phonon interactions. In many structures including double perovskites, cubic A$_2$BX$_6$, cubic perovskites, and KBrO$_3$, atoms with high $\mathcal{R}_{\text{onsite}}$ are located at the centers of cuboctahedra. Structures such as $\gamma$-LiFeO$_2$ and caswellsilverite exhibit high $\mathcal{R}_{\text{onsite}}$ at octahedrally coordinated cation sites, while trigonal A$_2$BX$_6$ families show high APRN effects in anticuboctahedral environments.

This analysis aligns well with previously reported experimental and theoretical studies that have identified strong anharmonic behavior in several classes of materials. Double perovskites and cubic perovskites are well-known for their significant anharmonic lattice vibrations~\cite{da2015phase, Yang2017, hoffman2023understanding, klarbring2020anharmonicity}. For example, Klarbring \textit{et al.}~\cite{klarbring2020anharmonicity} demonstrated that the lead-free halide double perovskite Cs$_2$AgBiBr$_6$ exhibits ultralow thermal conductivity, driven by strong lattice anharmonicity. Cubic A$_2$BX$_6$ compounds like Cs$_2$SnI$_6$ have also been shown to exhibit notable anharmonicity, where soft phonon modes are closely associated with octahedral tilting and structural instability~\cite{jong2019anharmonic, bhumla2022vacancy}. In the case of pyrochlore iridates, several groups have investigated phonon anomalies using Raman spectroscopy~\cite{ueda2019phonon,kumar2023spin}. Ueda \textit{et al.} observed pronounced phonon softening and line-shape anomalies in the Ir–O–Ir bond-bending vibrations of Eu$_2$Ir$_2$O$7$, indicating strong electron–phonon interactions. Similarly, Kumar \textit{et al.} identified strong spin–phonon and electron–phonon coupling in (Y$_{1-x}$Pr$_x$)$_2$Ir$_2$O$_7$, further enriching the temperature-dependent lattice dynamics in these materials.

We have so far used the fine-tuned M3GNet model to efficiently screen materials that exhibit strong APRN effects. To validate this data-driven approach, we carried out first-principles calculations based on DFT combined with SCPH theory. It is expected that strong APRN leads to shifts in phonon frequencies and reductions in phonon lifetimes, which directly impact $\kappa_L$. To systematically evaluate the impact of TDPH on $\kappa_L$, we selected 12 representative materials with high predicted $\mathcal{R}_{\text{onsite}}$ values. For each material, $\kappa_L$ was computed under three theoretical approximation: (i) harmonic approximation with 3-phonon scattering (HA + 3ph), (ii) SCPH approximation with 3-phonon scattering (SCPH + 3ph), and (iii) SCPH with both 3- and 4-phonon scattering (SCPH + 3,4ph), as adopted in our earlier study~\cite{rczbprx}. The calculated $\kappa_L$ values for these materials are presented in Fig.~\ref{fig:prototypes}D. When comparing HA + 3ph (gray bars) and SCPH + 3ph (blue bars), we observe a consistent increase in $\kappa_L$ for all materials. However, incorporating 4-phonon scattering processes (SCPH + 3,4ph, red bars) notably suppresses $\kappa_L$. Among the studied materials, Cs$_2$RbGaF$_6$, a double perovskite with significant variation in thermal conductivity across the different theoretical approximations, was chosen for more detailed analysis. The computed $\kappa_L$ values for Cs$_2$RbGaF$_6$ were 0.61, 2.15, and 0.98 W/(m·K) under HA + 3ph, SCPH + 3ph, and SCPH + 3,4ph, respectively.

To understand these variations, we analyzed the underlying phonon properties of Cs$_2$RbGaF$_6$, as shown in Fig.~\ref{fig:prototypes}E-F. The corresponding analyses for the remaining 11 materials are presented in the Supplementary Material. Fig.~\ref{fig:prototypes}E presents its phonon dispersion and projected density of states (pDOS) of Cs$_2$RbGaF$_6$ at 0~K (black) and 300~K (red). Consistent with the predictions from the fine-tuned M3GNet model, our SCPH calculations confirm significant APRN at 300~K, particularly evident in the low-frequency region. The pDOS indicates that F$^-$ ions dominantly contribute to vibrational modes throughout the entire frequency range, with particularly strong contributions in the acoustic region, where significant phonon renormalization occurs. To clarify the origin of $\kappa_L$ changes, Fig.~\ref{fig:prototypes}F compares phonon scattering rates ($\tau^{-1}$) obtained from each theoretical approach. Incorporating APRN via SCPH reduces scattering rates, especially in the low-frequency region (approximately 0-4 THz). This decrease in scattering rates increases phonon lifetimes ($\tau$), thus enhancing $\kappa_L$. However, when four-phonon scattering is included, scattering rates increase again, reducing phonon lifetimes and suppressing $\kappa_L$. This result shows the general cancellation of errors and highlights the significant contribution of four-phonon scattering even at room temperature. Our results highlight the necessity to fully account for higher-order anharmonicity in materials exhibiting strong TDPH.

%---------------------------------------------------------
%---------------------------------------------------------
\subsection{Future directions and framework extensions}

Before closing, we would like to point out several promising direction to advance our understanding of TDPH beyond the current framework. The rapid development of foundation models in materials science presents immediate opportunities to incorporate more recent and accurate machine learning interatomic potentials, potentially improving both harmonic and anharmonic property predictions. Our current approach focuses primarily on phonon frequency shifts due to anharmonic renormalization, which naturally leads to extending the framework to include phonon linewidth broadening and finite lifetimes for a complete description of temperature-dependent behavior. Achieving full self-consistency in our SSCHA calculations represents another important advancement that would improve accuracy for materials with extreme anharmonic effects, though this enhancement must be carefully balanced against computational efficiency requirements for high-throughput screening applications. Expanding our analysis beyond 300~K to cover wider temperature ranges will enable prediction of high-temperature material behavior relevant for energy applications and provide insights into temperature-dependent phase transitions. Finally, developing direct machine learning models to predict anharmonic phonon properties without explicit lattice dynamics calculations could further accelerate materials screening while maintaining acceptable accuracy for identifying qualitative trends and promising candidate materials. We envision that our approach can be used to discover materials relevant to applications in superionic conductors, thermal barrier coatings, thermoelectrics, and solid-state electrolytes, where anharmonic effects often play a defining role in transport properties and structural stability.

%---------------------------------------------------------
%---------------------------------------------------------
\section{Conclusion}
In summary, we have developed a computational framework for data-driven understanding of temperature-dependent phonons across diverse materials. By combining machine learning interatomic potentials with efficient anharmonic lattice dynamics, we demonstrate the feasibility of large-scale screening of temperature-dependent phonons for thousands of materials, previously computationally prohibitive using traditional approaches. Our key contributions include developing a fine-tuned M3GNet model with dramatically improved accuracy (reducing errors from 1.09 THz to 0.28 THz), implementing a streamlined SSCHA variant that enables high-throughput computation while maintaining accuracy, and systematically analyzing 4,669 compounds to reveal clear elemental and structural trends in anharmonic behavior. Moreover, machine learning identification of critical atomic environments underlying strong temperature effects provides valuable design guidelines, showing that alkali metals and perovskite-related structures with weakly bonded atoms in oversized coordination environments exhibit pronounced anharmonicity. First-principles validation demonstrates significant impacts on thermal conductivity, with temperature effects altering transport properties by factors of 2-4, highlighting the importance of realistic temperature considerations in materials design. Our framework promises a systematic exploration of anharmonic effects for thermal management, phase transition prediction, and phonon-mediated property design, establishing a foundation for next-generation materials engineering under finite temperatures.

%---------------------------------------------------------
%---------------------------------------------------------
\section{Methods}

\subsection{M3GNet fine-tuning}
The fine-tuning procedure utilized displacement configurations generated from harmonic interatomic force constants (IFCs) extracted from the MDR database~\cite{togo_database}, with ten supercell configurations created per compound using random small atomic displacements consistent with the harmonic approximation (0.01 - 0.05~\r{A}). This method ensures proper sampling of diverse atomic environments and vibrational modes while maintaining consistency with the underlying harmonic phonon calculations in the database. The resulting dataset was randomly split into training (95\%) and validation/test (5\%) sets to ensure robust model evaluation and prevent overfitting. The expanded training dataset was used to fine-tune the pre-trained M3GNet model over 150 epochs using a loss function based on the mean absolute error (MAE) for forces. The training process emphasized optimization of force predictions, considering that phonon predictions rely exclusively on accurate force calculations. Final training convergence was achieved with a force MAE of 0.0269 eV/\r{A} for both the training and validation/test sets, demonstrating excellent generalization performance. The fine-tuning hyperparameters (e.g., learning rates) were systematically optimized to maintain the model's transferability while specifically enhancing its performance on force predictions. The learning curve showing the validation/test MAE of forces during M3GNet fine-tuning is provided in Supplementary Material Fig. S1.

\subsection{Random Forest model} To ensure model reliability, we applied an outlier filtering process before constructing random forest models. First, we compared the DFT-calculated phonon frequencies with those predicted by our fine-tuned M3GNet model. Samples with prediction errors larger than 20\% were excluded. Second, we filtered out samples where the relative change in phonon frequencies predicted by the fine-tuned M3GNet model before and after APRN at 300~K exceeded 40\%. The thresholds were chosen based on visual inspection of the phonon dispersion curves. The features used for the random forest model training were extracted using the Matminer package~\cite{ward2018matminer}. For the global APRN model, we used 135 composition- and structure-based features, while the local APRN model was trained using 67 local features. A complete list of features used is provided in the Supplementary Material. For both models, the dataset was randomly split into 90\% training and 10\% testing sets. Hyperparameter tuning was performed using a grid search, with details provided in the Supplementary Material. The global APRN model was trained with n\_estimators = 200, max\_depth = 18, and max\_features = 0.5. The local APRN model used n\_estimators = 200, max\_depth = 15, and max\_features = 0.7. Due to the skewed distribution of $\mathcal{R}_{\text{onsite}}$, we applied a Box–Cox transformation with a lambda value of 0.074 to the target variable prior to training.

\subsection{First-principles calculation of lattice thermal conductivity} We conducted first-principles simulations to calculate the vibrational and thermal transport properties of the 12 selected materials in the main text using the Vienna Ab initio Simulation Package (VASP)~\cite{Vasp1, Vasp2, Vasp3, Vasp4}. Our computational approach encompassed structure relaxation and self-consistent density functional theory (DFT) calculations. We employed the projector-augmented wave (PAW) method \cite{PAW} in combination with the Perdew-Burke-Ernzerhof revised for solids (PBEsol)\cite{PBE,Perdew2008} generalized gradient approximation (GGA)\cite{GGA} for the exchange-correlation (xc) functional.\cite{DFT}.
For both structure relaxations and self-consistent calculations, we utilized Gamma-centered $\mathbf{k}$-point meshes with a KSPACING value of 0.2. Throughout the study, we maintained a kinetic energy cutoff of 520 eV. To ensure computational accuracy, we set stringent convergence criteria: $10^{-3}$ eV/\r{A} and $10^{-8}$ eV for force and energy, respectively.
To compute harmonic phonon dispersion, we employed the Phonopy software package \cite{phonopy} and used supercell sizes consistent with the MDR phonon database~\cite{togo_database}. Compressive sensing lattice dynamics (CSLD)\cite{csld,csld1} was utilized to extract anharmonic IFCs, with the cutoff distances for 3rd- and 4th-order IFCs limited up to 4th and 2nd nearest neighbours, respectively. These cutoff distances have been verified to achieve good convergence of lattice thermal conductivity, which was calculated using our own implementation of self-consistent phonon theory~\cite{xia2018revisiting} and three- and four-phonon scatterings~\cite{rczbprx}.

%---------------------------------------------------------
%---------------------------------------------------------

%%%%%%%%%%%%%%%%%%%%%%%%%%%%%%%%%%%%%%%%%%%%%%%%%%%%%%%%%%%%%%%%%%%%%
%% The "Acknowledgement" section can be given in all manuscript
%% classes.  This should be given within the "acknowledgement"
%% environment, which will make the correct section or running title.
%%%%%%%%%%%%%%%%%%%%%%%%%%%%%%%%%%%%%%%%%%%%%%%%%%%%%%%%%%%%%%%%%%%%%
\begin{acknowledgments}

H. L. and Y. X. acknowledge support from the U.S. National Science Foundation through award DMR-2317008. Y.X. also acknowledges support from the U.S. National Science Foundation through award CBET-2445361. We acknowledge the computing resources provided by Bridges2 at Pittsburgh Supercomputing Center (PSC) through allocations mat220006p and mat220008p from the Advanced Cyber-infrastructure Coordination Ecosystem: Services \& Support (ACCESS) program, which is supported by National Science Foundation grants \#2138259, \#2138286, \#2138307, \#2137603, and \#2138296.

\end{acknowledgments}

%%%%%%%%%%%%%%%%%%%%%%%%%%%%%%%%%%%%%%%%%%%%%%%%%%%%%%%%%%%%%%%%%%%%%
%% The same is true for Supporting Information, which should use the
%% suppinfo environment.
%%%%%%%%%%%%%%%%%%%%%%%%%%%%%%%%%%%%%%%%%%%%%%%%%%%%%%%%%%%%%%%%%%%%%
Supplementary material contains details on the M3GNet  fine-tuning, first-principles validation of 11 down-selected materials, random forest machine learning models, and a list of materials identified with strong anharmonic phonon renormalization.
The codes, data sets, and machine learning models will be made available via public repository upon the acceptance of the manuscript. 

%%%%%%%%%%%%%%%%%%%%%%%%%%%%%%%%%%%%%%%%%%%%%%%%%%%%%%%%%%%%%%%%%%%%%
%% The appropriate \bibliography command should be placed here.
%% Notice that the class file automatically sets \bibliographystyle
%% and also names the section correctly.
%%%%%%%%%%%%%%%%%%%%%%%%%%%%%%%%%%%%%%%%%%%%%%%%%%%%%%%%%%%%%%%%%%%%%
%\bibliography{achemso-demo}
\bibliography{minikappa}

\end{document}